\documentclass[reprint,superscriptaddress,amsmath,amssymb,aps]{revtex4-1}  
\usepackage{graphicx}
\usepackage{dcolumn}
\usepackage{bm}
\usepackage{color}
\usepackage{epstopdf} 
\usepackage{enumerate}
\usepackage{braket}
\usepackage{siunitx}

\begin{document}
\title{Phonon Magnetochiral Effect}
\author{T. Nomura}
\thanks{These two authors contributed equally}
\affiliation{Hochfeld-Magnetlabor Dresden (HLD-EMFL), Helmholtz-Zentrum Dresden-Rossendorf, 01314 Dresden, Germany}
\author{X.-X. Zhang}
\thanks{These two authors contributed equally}
\affiliation{Department of Applied Physics, The University of Tokyo, Tokyo 113-8656, Japan}
\affiliation{Quantum Matter Institute, University of British Columbia, Vancouver BC, V6T 1Z4, Canada}
\author{S. Zherlitsyn}
\affiliation{Hochfeld-Magnetlabor Dresden (HLD-EMFL), Helmholtz-Zentrum Dresden-Rossendorf, 01314 Dresden, Germany}
\author{J. Wosnitza}
\affiliation{Hochfeld-Magnetlabor Dresden (HLD-EMFL), Helmholtz-Zentrum Dresden-Rossendorf, 01314 Dresden, Germany}
\affiliation{Institut f\"ur Festk\"orper- und Materialphysik, TU-Dresden, 01187 Dresden, Germany}
\author{Y. Tokura}
\affiliation{Department of Applied Physics, The University of Tokyo, Tokyo 113-8656, Japan}
\affiliation{RIKEN Center for Emergent Matter Science (CEMS), Wako 351-0198, Japan}
\author{N. Nagaosa}
\affiliation{Department of Applied Physics, The University of Tokyo, Tokyo 113-8656, Japan}
\affiliation{RIKEN Center for Emergent Matter Science (CEMS), Wako 351-0198, Japan}
\author{S. Seki} 
\affiliation{RIKEN Center for Emergent Matter Science (CEMS), Wako 351-0198, Japan}

\date{\today}

\begin{abstract}
Magnetochiral effect (MChE) of phonons, a nonreciprocal acoustic property arising due to the symmetry principles, is demonstrated in a chiral-lattice ferrimagnet Cu$_2$OSeO$_3$.
Our high-resolution ultrasound experiments reveal that the sound velocity differs for parallel and antiparallel propagation with respect to the external magnetic field.
The sign of the nonreciprocity depends on the chirality of the crystal in accordance with the selection rule of the MChE.
The nonreciprocity is enhanced below the magnetic ordering temperature and at higher ultrasound frequencies, which is quantitatively explained by a proposed magnon-phonon hybridization mechanism. 
\end{abstract}

\maketitle 

There are physical phenomena which arise due to the fundamental symmetry principles; mirror symmetry breaking in chiral matters leads to natural optical/acoustic activity and time-reversal symmetry breaking by magnetic fields leads to magnetic optical/acoustic activity.
When both symmetries are simultaneously broken, a nonreciprocal property appears, the so called magnetochiral effect (MChE). 
The MChE has been observed for photons \cite{Rikken97,Vallet01,Koerdt03,Train08,Okamura15}, electrons \cite{Rikken01,Krstic02,Pop14,Yokouchi17}, and magnons \cite{Iguchi15,Seki16,Takagi17}.
Here, the (quasi)particles with the propagation vector $\bf k$ parallel and antiparallel to the magnetic field $\bf H$ show different propagation properties.
Owing to the symmetry origin of the MChE (time-reversal and mirror symmetry breakings), any changes in the sign of $\bf H$ or the chirality of the crystals ($\sigma=\pm1$) result in a reversed nonreciprocity for $\pm \bf k$.

From the symmetry consideration, the MChE should exist also for phonons (elementary vibrations of a crystal lattice) \cite{Szaller13}, which has not been observed experimentally. 
A phonon is one of the most ubiquitous quasiparticle, responsible for the heat and sound transmissions in solids.
The nonreciprocal phonons could enable the rectifications of the heat and sound, which are attractive for phononics applications \cite{Li12,Maldovan13,Connell10,Fornieri17}.
However, nonreciprocal flows of heat and sound have been realized only in artificial structures \cite{Chang06,Liang10,Fleury14,Walker18} or for surface-acoustic waves \cite{Heil82,Sasaki17}.
The MChE of phonons, which realizes the nonreciprocal thermal and acoustic flows in bulk media, is a promising strategy for future technologies.
Furthermore, confirming the universality of the MChE for the representative (quasi)particles (photon, electron, magnon, and phonon) will be an important milestone for fundamental science. 
Herein, we have revealed the last missing piece of them, the phonon MChE, with high-resolution ultrasound measurements on Cu$_2$OSeO$_3$.

Our target compound Cu$_2$OSeO$_3$ possesses a chiral cubic crystal structure with space group $P2_13$, as shown in Fig. 1. Both left-handed (L, $\sigma=-1$) and right-handed (D, $\sigma=+1$) single crystals can be obtained, which are distinguishable by natural optical activity measurements. This compound contains two inequivalent Cu$^{2+}$ ($S=1/2$) sites with a ratio of 3:1. 
As a result, a local ferrimagnetic spin arrangement with three-up-one-down manner is stabilized below $T_\mathrm{C}\approx58$ K \cite{Bos08}. 
The Dzyaloshinskii-Moriya (DM) interaction causes an additional long-period ($\lambda\approx 62$ nm) helical spin modulation at $H=0$, and external magnetic fields induce successive magnetic phase transitions from helical to conical and to collinear magnetic configurations \cite{Seki12PRB_neutron}. 
Because of the interplay between the chiral crystal symmetry and magnetism, various exciting properties such as magnetic skyrmions \cite{Seki12Science,White14,Seki12PRB_neutron}, multiferroicity \cite{Seki12Science,White14,Bos08,Seki12PRB}, and magnon MChE \cite{Seki16} have been reported.
Therefore, Cu$_2$OSeO$_3$ is an ideal candidate to observe the phonon MChE.

The high-resolution sound-velocity measurements are based on the ultrasonic pulse-echo technique with a phase-sensitive detection.
Two LiNbO$_3$ transducers were attached on the polished surfaces of the homochiral single crystals of Cu$_2$OSeO$_3$ for excitation/detection.
The relative changes of the sound velocity $\Delta v/v_0$ for L- and D-type  crystals ($\sigma=\pm1$) were measured with $\pm \bf{k}$ and $\pm \bf{H}$ geometries to examine the MChE (Fig. 1).
The results are compared for the different acoustic modes, ultrasound frequencies ($\omega/2\pi$), and temperatures ($T$).
For details, see Supplemental Material (SM) \cite{Supple}.

\begin{figure}[tb]
\centering
\includegraphics[width=0.85\linewidth]{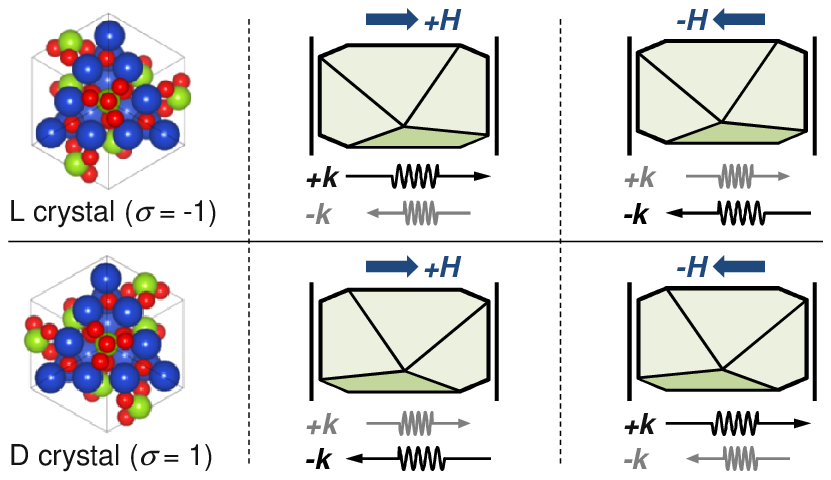}
\caption{
Experimental configurations to examine the phonon MChE.
Chirality of the crystal ($\sigma=\pm1$), sound propagation direction ($\pm \bf{k}$), and magnetic field direction ($\pm \bf{H}$) are successively inverted and the relative sound velocities are compared. 
The L- and D-type crystal structures of Cu$_2$OSeO$_3$ are shown along the [111] direction, where the spheres represent Cu (blue), O (red), and Se (green) atoms.
} 
\end{figure}

Figure 2(a) shows the results for the transverse acoustic (TA, with displacement vector ${\bf u} \perp {\bf k}$) mode in the L crystal measured in Faraday geometry (${\bf H} \parallel {\bf k}$) at 2 K.
Magnetic phase transitions from helical to conical ($H_{c1}=30$ mT) and from conical to collinear ($H_{c2}=95$ mT) states are observed as a step and a minimum in $\Delta v/v_0$, respectively (see also Ref. \cite{Evans17}). 
Remarkably, $\Delta v/v_0$ slightly differs at $H_{c2}$ for $\pm \bf{k}$, that is a clear feature of the nonreciprocity.
When the magnetic field is reversed, the sign of the observed nonreciprocity becomes opposite.
For a crystal with opposite chirality (D-type, Fig. 2(b)), all these nonreciprocal relations are reversed.
Therefore, the nonreciprocity is reversed for any changes in sign of $\sigma$, $\bf{k}$, and $\bf{H}$.
Such a nonreciprocity disappears in Voigt geometry ($\bf{k} \perp \bf{H}$) for the same TA mode (Fig. S2 in SM \cite{Supple}).
These selection rules evidence that the observed nonreciprocity originates from the phonon MChE.
Interestingly, the magnitude of the MChE for the longitudinal acoustic (LA, with ${\bf u} \parallel {\bf k} \parallel {\bf H}$) mode is much smaller (Fig. 2(c))  and cannot be identified within the experimental resolution of $\Delta v/v_0 \approx 10^{-6}$.

\begin{figure}[tb]
\centering
\includegraphics[width=0.92\linewidth]{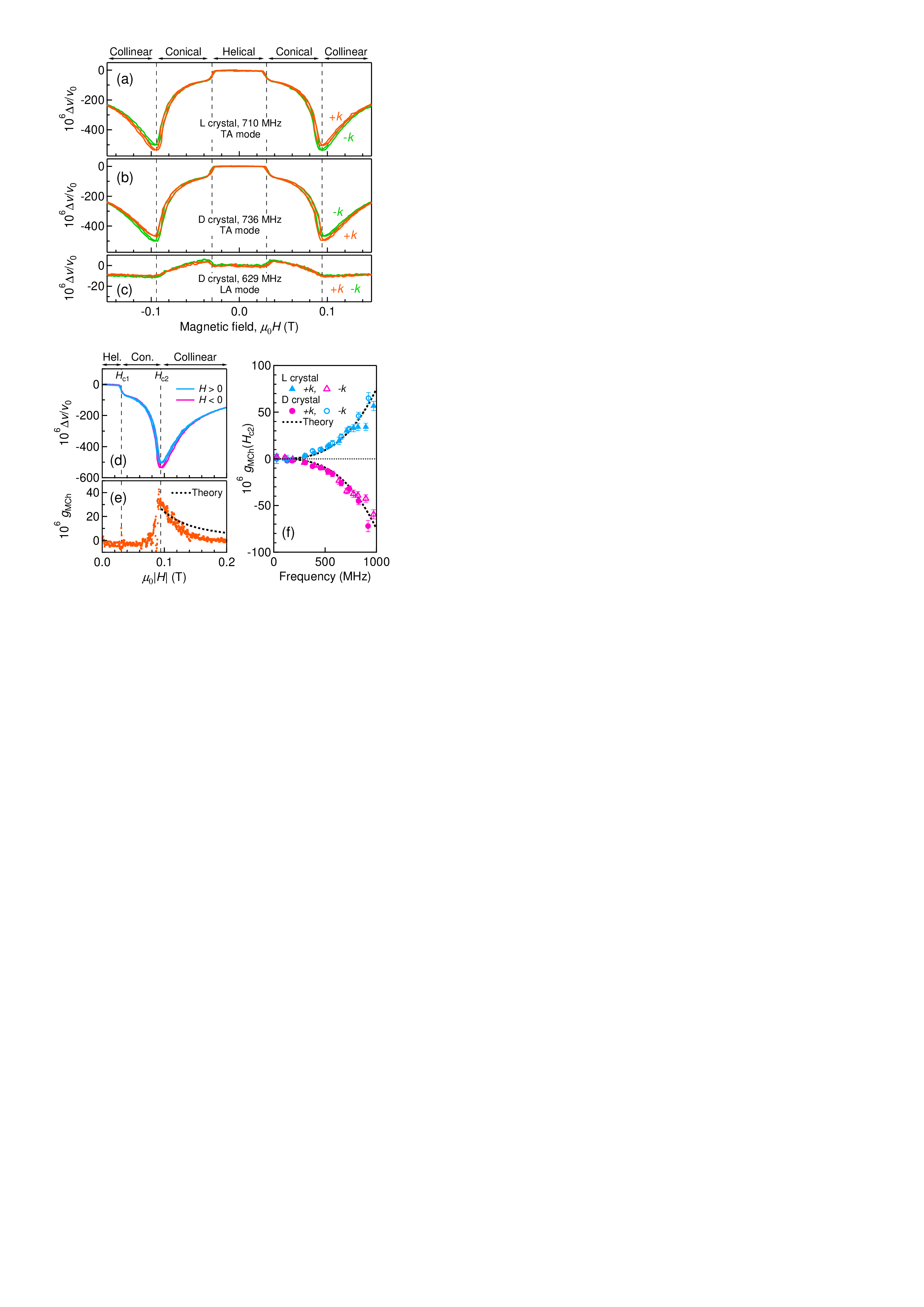}
\caption{
(a)--(c) Relative changes of the sound velocity $\Delta v/v_0$ as a function of magnetic field $H$ at 2 K. 
The transverse $(c_{11}-c_{12})/2$ mode (${\bf k}\parallel{\bf H}\parallel[110]$, ${\bf u}\parallel[1\overline{1}0]$) for the (a) L and (b) D crystal, and (c) the longitudinal $(c_{11}+c_{12}+2c_{44})/2$ mode (${\bf k}\parallel{\bf u}\parallel{\bf H}\parallel[110]$) for the D crystal are presented.
The sound propagation direction ($\pm \bf{k}$) and the ultrasound frequency are denoted for each curve.
All the experimental results are shown for up and down field sweeps. 
The sound velocities for the transverse and longitudinal modes are $v_{(c_{11}-c_{12})/2}=2.3$ km/s and $v_{(c_{11}+c_{12}+2c_{44})/2}=4.1$ km/s, respectively.
(d) $\Delta v/v_0$ as a function of $|H|$ for the transverse mode at 710 MHz with $+\bf{k}$ in the L crystal.
The results for $H>0$ ($H<0$) are shown by cyan (magenta). 
(e) Magnitude of the MChE $g_\mathrm{MCh}$ as a function of $|H|$, which corresponds to the difference between the data for $+H$ and $-H$ in Fig. 2(d).
(f) Maximum magnitude of the MChE $g_\mathrm{MCh}(H_\mathrm{c2})$ at 2 K as a function of ultrasound frequency for the transverse mode.
The results are shown for $\sigma = \pm 1$ and $\pm \bf{k}$.
The black dotted curves (Figs. 2(e) and 2(f)) denote the calculated results based on Eq. (\ref{eq:signal1}) with the magnetoelastic coupling constant $\gamma=90$.
} 
\end{figure}

For further analysis on the MChE of the TA mode, we plotted $\Delta v/v_0$ as a function of $|H|$ in Fig. 2(d).
Here, we introduce the degree of the MChE as
\begin{equation}
g_\mathrm{MCh}(H)=\frac{\Delta v(+{\bf H})}{v_0}-\frac{\Delta v(-{\bf H})}{v_0}=\frac{v(+{\bf H})-v(-{\bf H})}{v_0}.
\end{equation}
As shown in Fig. 2(e), the nonreciprocity is dramatically enhanced upon the transition into the collinear spin state ($H_{c2}$) and rapidly weakens at higher fields.
Figure 2(f) shows the maximum magnitude of the MChE $g_\mathrm{MCh}(H_\mathrm{c2})$ as a function of ultrasound frequency.
The MChE is enhanced nonlinearly towards higher frequencies.

\begin{figure}[tb]
\centering
\includegraphics[width=0.99\linewidth]{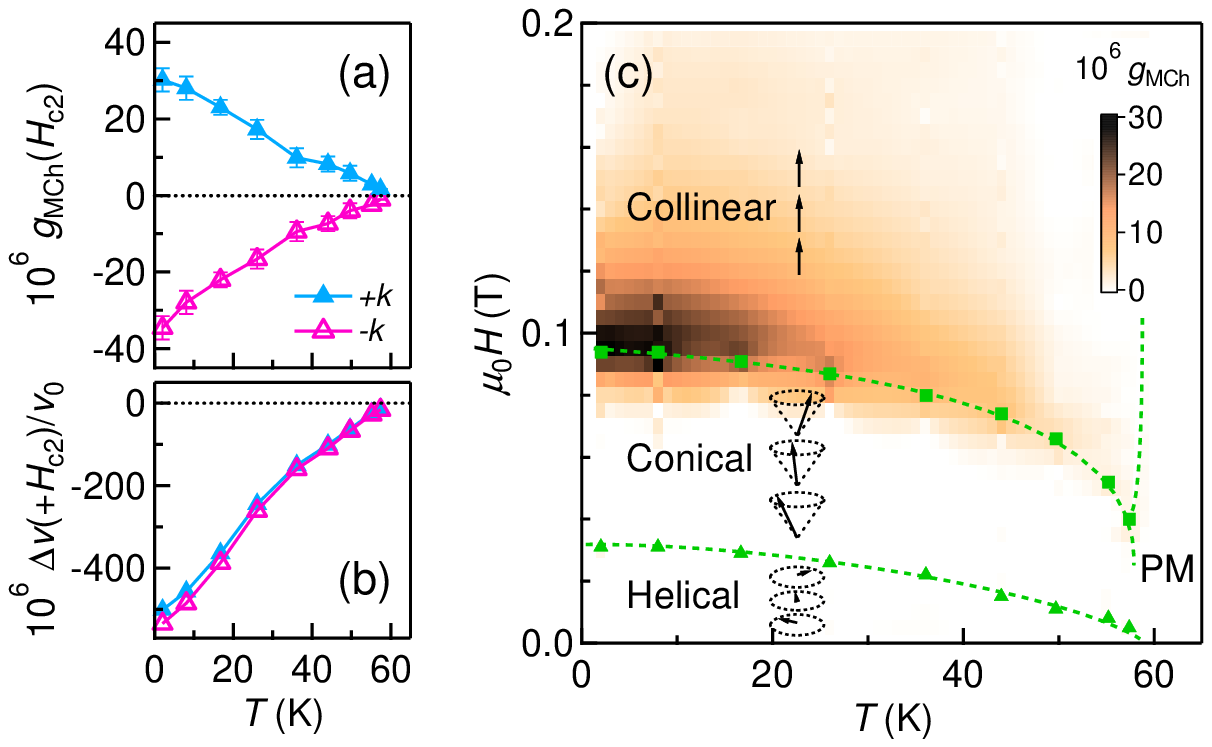}
\caption{
Temperature dependences of (a) $g_\mathrm{MCh}(H_\mathrm{c2})$ and of (b) the relative change of the sound velocity at $H_\mathrm{c2}$ for the transverse mode.
(c) Contour plot of $g_\mathrm{MCh}$ mapped on the $T$-$H$ phase diagram.
The phase boundaries are determined by the anomalies in $\Delta v/v_0$ (triangles and squares).
Here, the results for the L crystal at 710 MHz are used.
} 
\end{figure}

$T$ dependence of $g_\mathrm{MCh}(H_\mathrm{c2})$ is shown in Fig. 3(a).
The MChE weakens at higher $T$ and linearly vanishes with approaching $T_\mathrm{C} \approx 58$ K.
With increasing $T$, the relative change of the sound velocity itself also decreases (Fig. 3(b)).
For an overview, the contour plot of $g_\mathrm{MCh}$ is mapped on the $T$-$H$ phase diagram (Fig. 3(c)).
Obviously, the presence of nonreciprocity is related to the underlying spin structure. The MChE is absent in the helical and conical spin states, and $g_\mathrm{MCh}$ takes maximum at near $H_\mathrm{c2}$ in the collinear spin state.

When considering the mechanism of the phonon MChE, there are three important features.
First, the MChE appears at $H_{c2}$ and rapidly weakens at higher fields (Fig. 2(e)).
The absence of the MChE below $H_{c2}$ (helical and conical states) is observed also for the magnons in Cu$_2$OSeO$_3$, where the DM interaction causes the asymmetric magnon dispersion in the collinear spin state but such an asymmetry between $\pm {\bf k}$ is lost by the folding back of magnon branch in the helical/conical spin states \cite{Seki16,Kataoka87}.
This common feature indicates a correlation between the MChEs of magnons and phonons.
Second, the MChE is dramatically enhanced at higher ultrasound frequencies (Fig. 2(f)).
In our experiment, the phase velocity of acoustic phonons $v=\omega/k$ is measured.
Ultrasound frequencies of a few hundreds MHz are rather close to the Brillouin-zone center where the acoustic phonon dispersion is usually linear.
The observed frequency dependence indicates that the MChE is related to the nonlinear dispersion of the acoustic phonons.
Third, the magnitude of the MChE and $\Delta v/v_0$ for the LA mode are much smaller than the TA mode (Figs. 2(a)--(c)).
This implies that the shear strain connected with the TA phonon plays a dominant role in the observed MChE.

The underlying picture is that the acoustic phonons inherit the nonreciprocity from the asymmetric magnon excitations via a magnon-phonon band hybridization (Fig. 4(b)).
This magnon-phonon hybridization results in a band repulsion or anticrossing, which deforms the linear dispersion of the TA phonons.
When the magnon dispersion possesses an asymmetry due to the DM interaction \cite{Seki16,Kataoka87}, the hybridization points differ for $\pm{\bf{k}}$ and the acoustic phonons acquire the nonreciprocity.
When the ultrasound frequency is close to the hybridization points, the MChE is dramatically enhanced.
Besides, the magnitude of the MChE depends on the degree of the anticrossing produced by the magnetoelastic coupling.

\begin{figure}[tb]
\centering
\includegraphics[width=0.95\linewidth]{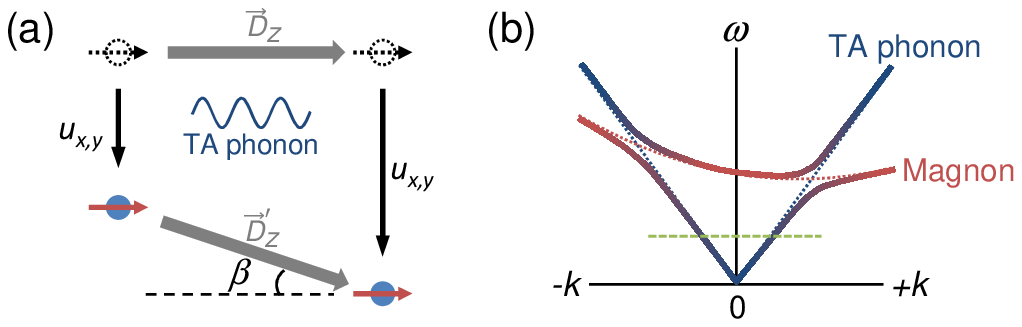}
\caption{
(a) Magnetoelastic coupling due to the shear strain. 
TA phonons cause the displacements $u_{x,y}$ of the neighboring lattice sites (blue circles) from the original places (dotted circles). The original DM vector $\vec{D}_z$ is modified to $\vec{D}_z'$ by a tilting angle $\beta$.
The magnetic moments (dashed and red arrows) and the TA phonon propagation are along the $\hat{z}$ direction. In reality, the two displacements are not necessarily in the same direction.
(b) Dispersions of the magnon and phonon bands hybridized by the chiral magnetoelastic coupling. 
The original dispersions are shown by dotted lines. 
One of the circularly polarized phonon mode hybridizes with the magnons, leading to the anticrossing. Another mode does not hybridize and its dispersion remains as the original blue dotted line. 
The ultrasound frequency used in this study (dashed line) was always lower than the hybridization frequency.
} 
\end{figure}

For the origin of the magnon-phonon hybridization, we propose a chiral magnetoelastic coupling due to the modulation of the DM interaction by shear strains.
The major effect of TA phonons is to tilt the bonds and to modify the DM interaction, in contrast to the elongation/compression due to LA phonons \cite{Zhang17,Kanazawa16}.
We base the discussion on the minimal magnetic Hamiltonian\\
 $\mathcal{H}_\mathrm{latt}= \sum_{\vec{r},\hat{\alpha}} { [ -J  \vec{S}_{\vec{r}} \cdot\vec{S}_{\vec{r}+\hat{\alpha}} -\vec{D}_{\alpha} \cdot \vec{S}_{\vec{r}} \times \vec{S}_{\vec{r}+\hat{\alpha}}  - g \vec{S}_{\vec{r}}\cdot\vec{B} ]}$, including the nearest-neighbor exchange interaction $J$, the DM interaction $D$, and the Zeeman energy, in which $g=2\mu_\mathrm{B}$ and $\vec{D}_\alpha$ denotes the DM vector on the bond along $\hat{\alpha}$ direction.
The collinear phase is treated as an effective ferromagnetic state averaged over the unit cell. 
We take into account spin waves, i.e., fluctuations $\delta \vec{S}$ in this ordered phase, coupled with phonons propagating along the direction $\hat{z} \parallel \bf{H}$.
Crucially, the leading-order effect should result from the harmonic magnetoelastic coupling, i.e., bilinear in both the spin-wave and phonon coordinates. 
As can be seen from the form of $\mathcal{H}_\mathrm{latt}$, the desired coupling must come from the DM interaction because the spin-wave fluctuations of the exchange term are always quadratic in $\delta\vec{S}\perp\hat{z}$ in this state with ordered moment $\braket{S^z}$. 
Most importantly, for the acoustic wave propagating parallel to $\hat{z}$, only the TA phonon can generate bilinear couplings through modifying $\vec{D}_z$ while the LA phonon leads to higher order effects, e.g., two-magnon processes due to couplings quadratic in the transverse spin moments. This favourably explains the observed (ir)relevance of TA (LA) phonons.

As shown in Fig. 4(a), the TA phonon displacement $\vec{u}=(u_x,u_y)$ changes the DM vector to $\vec{D}_z' = D\hat{z} + \gamma D \, \partial_z u_i \, \hat{i}$ by a small angle $\beta=\partial_z |\vec{u}|$.
Here, we introduce a phenomenological dimensionless magnetoelastic coupling constant $\gamma$ since the tilting induced by TA phonons is a lattice shear strain rather than merely a rotation of the bonds. Then, one can arrive at the Lagrangian density describing the chiral bilinear interaction
$\mathcal{L}_\mathrm{me}= \gamma D \braket{S^z}(-\partial_z u_x \partial_z S^y + \partial_z u_y \partial_z S^x)$.
This unique harmonic coupling proportional to the ordered moment $\braket{S^z}$ yields observable effects only quadratic in $\braket{S^z}$ as understood via a second-order perturbation. The ordered moment $\braket{S^z}$ and hence the experimentally detected responses are suppressed when $T$ increases. This particularly corroborates with the roughly linear $T$ dependences of both $g_\textrm{MCh}$ and $\Delta v/v_0(H_\mathrm{c2})$ as shown in Figs. 3(a) and 3(b) since $\braket{S^z}$ scales as $\sqrt{|T-T_\mathrm{C}|}$ near the magnetic phase transition.

This chiral coupling lifts the degeneracy in the pure phonon theory and makes two nondegenerate eigenstates of the left- and right-circularly polarized (LCP and RCP) TA phonon modes.
Here, only one phonon branch selectively hybridizes with magnons (Fig. 4(b)) since only right-handed polarization exists for ferromagnetic-type spin waves with respect to the magnetization, which is different from the case in antiferromagnets \cite{Lan17}. 
Experimentally, as a linearly polarized sound wave is injected and detected, a superposition of the intact and the hybridized circularly polarized waves results in an acoustic beat typically of the form $2\sin{(\frac{k^u+k^a}{2}z+\omega t)\cos{(\frac{k^u-k^a}{2}z)}}$, where $k^{a(u)}$ is the wave number (un)affected by the hybridization at the ultrasound frequency. 
Here, the beating envelope variation is negligibly small within the 2 mm sample thickness (see SM \cite{Supple}).
Therefore, one experimentally detects half the nonreciprocity that occurs for the hybridized phonons.

On the basis of these understandings, one can readily construct a solvable magnon-phonon interacting theory. 
Deriving an effective theory of the phonon, one is able to directly calculate the sound-velocity change
$\Delta v/v_0 = -\frac{ \gamma^2 \braket{S^z}^2 S D^2 k^2} {a_0 c \Delta_B}$
and the magnitude of the MChE
\begin{equation}\label{eq:signal1}
g_\mathrm{MCh} = \frac{ 4\gamma^2 \braket{S^z}^2 S^2 D^3 k^3} {c \Delta_B^2},
\end{equation}
in which $a_0,S,c,\Delta_B$ are the lattice constant, the total spin moment, the elastic constant, and the spin-wave gap, respectively (see SM for details \cite{Supple}).
This expression suggests that the MChE is enhanced at higher ultrasound frequencies ($k \propto \omega$) and at $H_\mathrm{c2}$ where the soft-mode magnon band hybridizes with the phonon band at rather low frequencies ($\Delta_B \approx$ 3 GHz) \cite{Seki16}.
The experimental results are compared with Eq. (2); the theoretically expected $g_\mathrm{MCh} \propto \omega^3$ relationship agrees well with the observed frequency dependence (Fig. 2(f)), and the $H$ dependence of $g_\mathrm{MCh}$ is also reproduced qualitatively (Fig. 2(e)).
The slight quantitative deviation might occur because the magnon-phonon hybridization point is too far from the Brillouin-zone center at higher $H$.
In the theoretical derivations, magnon and phonon dispersions are approximated to be linear in $k$, however at larger $k$, nonlinear dispersions become more relevant.
Nevertheless, the overall agreement between the theory and experiment, i.e. the significance of TA waves and reproduction of $T$-, $\omega$- and $H$-dependences of $g_\mathrm{MCh}$, strongly suggests that the present magnon-phonon hybridization mechanism associated with the nonreciprocal phonon dispersion well captures the essence of the observed phonon MChE.

In conclusion, the nonreciprocal sound propagations are observed in Cu$_2$OSeO$_3$, which are due to the phonon MChE.
The selection rule of the MChE is confirmed by the complete set of experimental geometries.
The magnitude of the MChE as a function of $\omega$, $H$, and $T$ is well explained by the proposed magnon-phonon hybridization mechanism.
While the observed magnitude of MChE remains rather small at this stage, its $\omega^3$-dependence suggests that further enhancement of the effect is possible. 
In particular, when the excitation frequency is located within one of hybridization gaps, the unidirectional phonon propagation can be expected for the mode shown in Fig. 4(b). 
In principle, the phonon MChE can be expected for any chiral-lattice magnets, and the present results suggest a new intrinsic strategy for the nonreciprocal sound/heat transmission based on a single-phase bulk compound.

\bigskip
This work was partly supported by the JSPS Grants-In-Aid for Scientific Research (Nos.~18H03685, 17H05186, 16J07545, and 26103006), by the DFG through SFB 1143, and by HLD at HZDR, member of the European Magnetic Field Laboratory. T.N. was supported by the JSPS through a Grant-in-Aid for JSPS Fellows. N.N. was also supported by CREST, Japan Science and Technology (No.~JPMJCR16F1) and ImPACT Program of Council for Science, Technology and Innovation (Cabinet office, Government of Japan, 888176).

\end{document}